 \documentclass[final,5p,times,twocolumn]{elsarticle}

\usepackage{graphicx}
\usepackage{subfigure}

\usepackage[utf8]{inputenc}
\usepackage{amsmath}
\usepackage{amsthm}
\usepackage{amsfonts}
\usepackage{epsfig}
\usepackage{psfrag}
\usepackage{pstricks}
\usepackage{algorithm}
\graphicspath{{./Figures/}}

\usepackage{amssymb}
\usepackage{bm}

\def \div {\mathrm{div}}

 \biboptions{comma,sort&compress}

\journal{April 2020}

\begin{document}

\begin{frontmatter}



\title{Robust and efficient tool path generation for poor-quality triangular mesh surface machining}

\author[ubc]{Qiang Zou\corref{cor}}
\ead{john.qiangzou@gmail.com}

\cortext[cor]{Corresponding author.}
\address[ubc]{Department of Mechanical Engineering \\
The University of British Columbia \\
Vancouver, BC, Canada V6T 1Z4}

\begin{abstract}
This paper presents a new method to generate iso-scallop tool paths for triangular mesh surfaces. With the popularity of 3D scanning techniques, scanning-derived mesh surfaces have seen a significant increase in their application to machining. Quite often, such mesh surfaces exhibit defects such as noises, which differentiate them from the good-quality mesh surfaces previous research work focuses on. To generate tool paths for such poor-quality mesh surfaces, the primary challenge lies in robustness against the defects. In this work, a robust tool path generation method is proposed for poor-quality mesh surfaces. In addition to robustness, the method is quite efficient, providing the benefit of faster iterations and improved integration between scanning and machining. The fundamental principle of the method is to convert the tool path generation problem to the heat diffusion problem that has robust and efficient algorithms available. The effectiveness of the method will be demonstrated by a series of case studies and comparisons.

\end{abstract}


\end{frontmatter}


\section{Introduction}
\label{sec:introduction}
Tool path generation is a fundamental element in modern computer-aided design and manufacturing (CAD/CAM) systems as it bridges the geometry designed in CAD and the machining process controlled in CAM. The generated tool paths will govern how a machine tool moves its cutter relative to the workpiece during machining, and their engagement produces target geometries. Traditionally, these geometries are represented as parametric surfaces, and many effective tool path generation methods have been developed \cite{lasemi2010recent}. Recently, geometries represented as triangular mesh surfaces have seen increasing applications due to advances in 3D scanning techniques \cite{botsch2010polygon}. Although mesh surfaces could be approximately reconstructed to be parametric surfaces, many still prefer a direct, mesh-based tool path generation method, because the reconstruction remains a tedious task that requires significant expertise and user tuning to perform satisfactorily \cite{wen2017cutter}.

Direct tool path generation for mesh surfaces is no trivial matter. Quite often, mesh surfaces are constructed from scanned point cloud data and have a poor mesh quality. This issue may be alleviated using mesh repairing algorithms but cannot be avoided altogether. A direct tool path generation method should thus be robust against mesh defects. Although algorithms to directly generate tool paths for mesh surfaces have been reported \cite{xu2019contour}, application of them is consistently limited to good-quality mesh surfaces. In this paper, a robust, direct method to generate tool paths for three- or five-axis machining of mesh surfaces using ball-end mills is to be presented.  As mesh defects can take numerous forms in practice, a general method that can deal with all of them may not be possible, and a specialized scheme may be a more practical choice. In this regard, this work focuses on the following commonly found defects: noises, irregular density, and even small holes.

Besides the robustness requirement, high efficiency is another desirable feature in some applications. With the increasing popularity of 3D scanning, machining has seen a closer and closer integration with scanning, e.g., the systems developed in \cite{lasemi2012freeform, jones2012remanufacture}. As noted in these studies, fast iterations between scanning and machining are important to these systems. As an efficient tool path generation can lead to a reduced scanning-machining iteration time, it will be of significance to develop fast tool path generation methods. For this reason, this work also aims at efficient tool path generation for mesh surfaces, as a secondary feature of the proposed method.

The present work attains robustness and efficiency by establishing the connection between iso-scallop tool path generation and heat diffusion. Heat diffusion refers to a process where heat evolves over a surface of interest from a region of higher temperature to a region of lower temperature, or alternatively describes a collection of particles taking random walks starting from a given source region. As the mathematics behind random walks is local averaging, heat diffusion is robust to mesh defects, especially noises \cite{coifman2006diffusion}. Linear algorithms to simulate heat diffusion have also been made available \cite{crane2017heat}. As such, if the tool path generation problem can be formulated into the heat diffusion problem, a robust and efficient method to generate tool paths for poor-quality meshes can be developed. This states the very motivation of the present work. To the best of the author's knowledge, this conversion idea---from tool path generation to heat diffusion---has not been reported previously.

\section{Related work}
\label{sec:related-works}
Tool path generation methods can be roughly categorized into three strategies: iso-parametric, iso-planar and iso-scallop. Many mesh-based tool path generation methods also follow these strategies. The iso-parametric method generates tool paths for a parametric surface by keeping one of the surface's parameters constant and varying the other \cite{Qiang2013iso}. The parameter interval between adjacent tool paths is chosen in a way that the scallop height is constrained by a user-specified limit. This method was first adapted to mesh surfaces by Sun et al. \cite{yuwen2006iso}. Later, follow-up studies were proposed, with improvements on effectiveness \cite{hu2015boundary} and accuracy \cite{zhao2015tool}. Common to these methods is parametrization that reconstructs a parameter domain for the mesh surface. With this parameter domain, the iso-parametric strategy becomes readily applicable to mesh surfaces. This line of methods are computationally convenient, but the most serious limitation is the machining inefficiency caused by the redundant machining between two adjacent tool paths.

The iso-planar method was initially proposed as an improvement to the iso-parametric method. Its basic idea is to generate tool paths using the intersection curves between a series of parallel planes and the design surface \cite{huang1994non}. The distance between two adjacent planes is again chosen without exceeding the specified scallop height limit. This method has also been adapted to mesh surfaces by Yang and Feng \cite{yang2008machining} and Li et al. \cite{li2011tool}. The iso-planar method offers better control on the scallop height than the iso-parametric method; regions of dense tool paths observed in iso-parametric tool paths can be avoided considerably. Nevertheless, redundant machining still exists for iso-planar tool paths.

To avoid redundant machining completely, the iso-scallop method keeps the scallop between two adjacent tool paths the same as the specified limit. This method was first introduced by Suresh and Yang \cite{suresh1994constant} and improved in computational efficiency in \cite{koren1996efficient, yoon2005fast,li2004efficient} and accuracy in \cite{sarma1997geometry,feng2002constant,kim2007constant}. When spacing two adjacent tool paths with respect to the constant scallop height limit, both 2D and 3D algorithms can be used. The 2D algorithms carry out all calculations by reference to the 2D parameter domain of the surface, as in \cite{suresh1994constant}, while the 3D algorithms only use 3D surface-based calculations, as in \cite{sarma1997geometry}. The former has been successfully adapted to mesh surfaces by \cite{xu2013tool,xu2015mapping,xu2019contour} through mesh parameterization, and the latter by \cite{lee2008mesh,bolanos2014topological,kout2014tool,wen2017cutter}. This line of research was further developed in \cite{huertas2014obtaining,sun2016smooth,xu2018spiral,wang2018smooth} to generate tool paths that follow a spiral pattern. Although these methods are effective in many scenarios, no particular attention was paid to poor-quality mesh surfaces. As a result, their applicability to scanning-derived triangular mesh surfaces is limited. Surfaces of this kind, however, comprise an increasing and substantial portion of geometric models being used in CAD/CAM.

The above review suggests that, although considerable research efforts have been devoted to mesh-based tool path generation, existing methods will suffer when poor-quality meshes are used. To address this insufficiency, this paper presents a new method to robustly and efficiently generate iso-scallop tool paths for mesh surfaces with noise, irregular density, and even holes. It begins in the next section with recalling some fundamentals from machining and geometry, to be used as building blocks for the proposed method. Section 4 first outlines the basic ideas and then elaborates them with concrete formulations. Validation of the method using a series of tool path generation examples and comparisons is found in Section 5, followed by conclusions in Section 6.

\section{Machining geometry}
\label{section2}
To generate iso-scallop tool paths, effective representation and evaluation of the scallop height are necessary. We now recall some facts about scallop height calculation to support the presentation of the proposed method in the next section.

\begin{figure*}[htbp]
	\centering
	\includegraphics[]{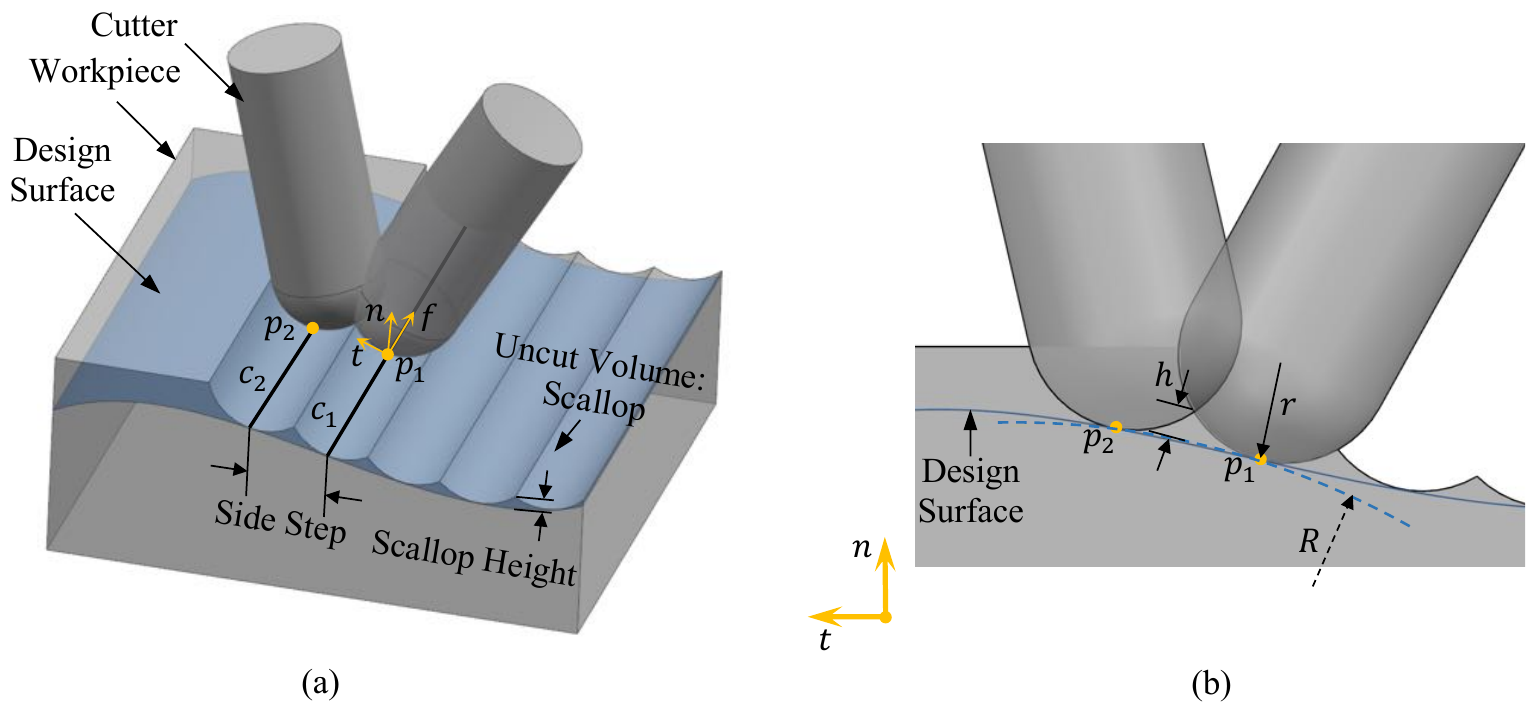}
    \caption{Illustration of scallop geometry and tool path parameters.}
	\label{path-parameters}
\end{figure*}

The scallop is the uncut volume left between a pair of adjacent tool passes due to the mismatch between the shape of the cutter and the shape of the design surface, as shown in Fig.~\ref{path-parameters}a. The scallop height refers to the maximum of the uncut volume's thickness, locally. This quantity is often evaluated with the help of the 2D sectional geometry shown in Fig.~\ref{path-parameters}b. The section plane is positioned at a cutter contact point $p_1$ and perpendicular to the feed direction $f$, and the 2D geometry is obtained by interesting the local design surface with the section plane and by projecting the cutter surface into this plane. In the case of ball-end milling, the projected geometry is a circular arc (if the cutter's shank is not considered). If other types of cutter are used, the projected geometry has various shapes \cite{lee1997surface}, but the concept of effective cutter radius can be used to carry out calculations as if they were circular arcs, see \cite{lo1999efficient} for a good example.

Based on the 2D sectional geometry, the scallop height can be easily calculated with a second-order approximation scheme: approximate the neighborhood of $p_1$ with its osculating circle with radius $R$. Then, the scallop height for a cutter with radius $r$ (or with effective cutter radius $r$) is given by \cite{koren1996efficient}:
\begin{equation}\label{eq-scallop-height}
	h=\frac{{R}+{r}}{8Rr}{{{\left\| p_2 - p_1 \right\|}^{2}}} + {O}\left( {{\left\| {{p}_{2}}-{{p}_{1}} \right\|}^{3}}\right),
\end{equation}
where $\left\| p_2 - p_1 \right\|$ is the distance between the two contact points, and is commonly referred to as the side step. In the above equation, $R$ is positive if the design surface is convex at $p_1$, and negative if concave.

The essential term in Eq.~\ref{eq-scallop-height} is $R$, which equals to the reciprocal of the normal curvature at $p_1$ in the tangential direction $t$ perpendicular to $f$. This normal curvature can be expressed in terms of the curvature tensor $T$ at $p_1$ as:
\begin{equation}\label{eq-normal-curvature}
	\kappa_t = v_t^{T} T v_t,
\end{equation}
with $v_t$ being the unit vector in direction $t$ \cite{Carmo76}. In the above expression, $T$ can be given by:
\begin{equation}\label{eq-curvature-tensor}
	T = \kappa_1 d_1 d_1^T + \kappa_2 d_2 d_2^T,
\end{equation}
where $\kappa_1, \kappa_2$ are the principal curvatures, and $d_1, d_2$ ($3 \times 1$ vectors) the principal directions. This way, $T$ is a symmetric $3 \times 3$ matrix with eigenvalues $\kappa_1, \kappa_2, 0$, and corresponding eigenvectors $d_1, d_2, n$, where $n$ is the normal vector. 

To calculate $\kappa_1, \kappa_2, d_1, d_2$ at a mesh vertex $v_i$, this work employs a local fitting method \cite{botsch2010polygon}: fit a quadric surface to a neighborhood of $v_i$ and then use this quadric surface to approximate curvatures. In this algorithm, the neighborhood size is user-specified, and it is set as vertices within $v_i$'s 2-ring neighborhood for all the tested models in this work. Eventually, Eq.~\ref{eq-curvature-tensor} evaluates to a $3 \times 3$ matrix at each vertex.  The per-triangle curvature tensor can be attained using an area-weighted average of the curvature tensors of incident vertices, following \cite{rusinkiewicz2004estimating}.

\section{The proposed methodology}
\label{sec:Methodology}

The method consists of two stages: (1) translate the iso-scallop tool path generation problem to the geodesic computation problem, and (2) convert the geodesic computation problem to the heat diffusion problem. The geodesic computation is to attain a geodesic function on the design surface that defines geodesic distances between source points (from which distance is measured) and other points on the surface. Altogether, these two stages yield a robust and efficient tool path generation method that involves only solving two sparse systems of linear equations.

\subsection{Translate tool path generation to geodesic computation}
\label{sec:scallop-metric}

\begin{figure}[tb]
	\centering
	\includegraphics[]{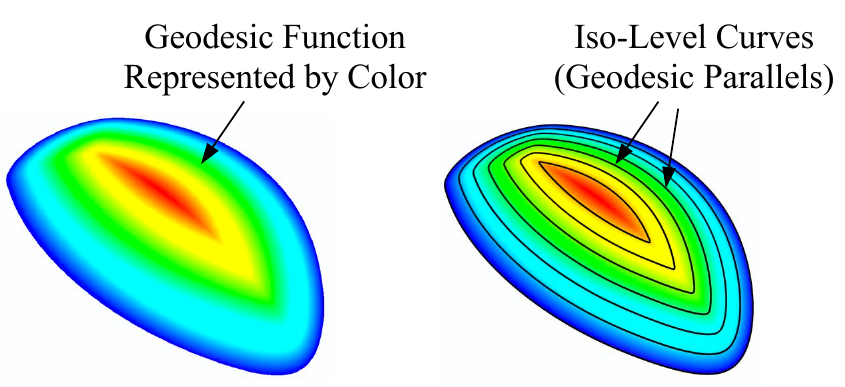}
    \caption{Example of the geodesic function and iso-level curves (geodesic parallels).}
	\label{geodesic-parallel}
\end{figure}

The translation to be presented basically says that a family of iso-scallop tool paths on a design surface can be made equivalent to a group of geodesic parallels on the same surface, but with a new Riemannian metric. The geodesic parallels refer to iso-level curves of the geodesic function on the surface, and an iso-level curve collects points on the surface that are mapped to the same value by the function \cite{zou2014iso}. Fig.~\ref{geodesic-parallel} shows an example of the geodesic function and several selected iso-level curves/geodesic parallels. The color variation represents the distance measured from the surface's boundary. 

The equivalence stated above can be shown with the following derivations. By substituting Eq.~\ref{eq-normal-curvature} into Eq.~\ref{eq-scallop-height}, we have:
\[
		h=\frac{v_t^{T} T v_t + \frac{1}{r}}{8}{{{\left\| p_2 - p_1 \right\|}^{2}}} + {O}\left( {{\left\| {{p}_{2}}-{{p}_{1}} \right\|}^{3}}\right).
	\]
The term $v_t^{T} T v_t \cdot {\left\| p_2 - p_1 \right\|}^{2}$ can be rewritten as:
\[
		v_t^{T} T v_t \cdot {\left\| p_2 - p_1 \right\|}^{2} = (p_2 - p_1)^T T (p_2 - p_1) + {O}\left( {{\left\| {{p}_{2}}-{{p}_{1}} \right\|}^{3}}\right).
\]
And the term $1/r \cdot {\left\| p_2 - p_1 \right\|}^{2}$ can be rewritten as:
\[
	\frac{1}{r} \cdot {\left\| p_2 - p_1 \right\|}^{2} = (p_2 - p_1)^T T_c (p_2 - p_1),
\]
where $T_c$ is a diagonal matrix with diagonal entries $1/r$. Then, the scallop height has the following new expression:
\begin{equation}\label{eq-scallop-curvature}
	h= (p_2 - p_1)^T {\frac{T + T_c}{8}} (p_2 - p_1) + {O}\left( {{\left\| {{p}_{2}}-{{p}_{1}} \right\|}^{3}}\right).
\end{equation}
Let $T_s = \frac{T + T_c}{8}$ and name it the scallop metric. This metric is to be used as a new Riemannian metric for the design surface\footnote{Matrices can be used to define a Riemannian metric as long as they are symmetric positive definite, and this is the case for $T_s$.}. Under this metric, distance between a pair of points $p, p + \Delta p$ is measured using:
\[
		d^2\left( p, p + \Delta p \right) = {\left( \Delta p \right)}^{T} T_s {\left( \Delta p \right)},
\]
instead of ${\left( \Delta p \right)}^{T} {\left( \Delta p \right)}$. As such, two curves having equal distance (e.g., $d$) regarding the scallop metric are two paths of constant scallop height ($d^2$).

At this point, generating iso-scallop tool paths is equivalent to attaining iso-level curves of the geodesic function. How to compute geodesic functions for mesh surfaces will be discussed in the next subsection. Before getting into it, a note should be made here. The insight of the equivalence between iso-scallop tool paths and geodesic parallels is not novel. Kim \cite{kim2007constant} first observed this equivalence in the parametric surface setting; Kout and Muller \cite{kout2014tool} then extends his results to mesh surfaces. Nevertheless, we derive a new, generalized scallop metric here, and this has a number of advantages including a much simpler expression for the metric without reference to any particular parametric domains or local coordinate systems (which are the cases in \cite{kim2007constant, kout2014tool}), improved computation efficiency, and most notably the generality coming from scallop heigh independence (the method in \cite{kout2014tool} is scallop heigh dependent and requires a metric construction for every new scallop height). 


\subsection{Convert geodesic computation to heat diffusion}
\label{sec:iso-scallop-diffusion}
As already noted, heat diffusion on a surface is essentially a random walking on the surface. During the walking, distance information will be collected in a way described by the Varadhan’s formula, which states that the geodesic distance $d$ between any pair of points $x, y$ on the surface is given by:
\[	
	d(x, y) = \lim_{t \to 0} \sqrt{-4t\ log\ h(t,y)}
\]
where $h(t,y)$ is the heat transferred from $x$ to $y$ at time $t$. This equation is of little usefulness in its own right as it is a limit; using a small time step to approximate the limit would not help, because the local averaging nature of heat diffusion blurs the calculated $d(x, y)$ \cite{coifman2006diffusion}. Fortunately, it has been found by Crane et al. \cite{crane2017heat} that, although $d(x, y)$ is blurred under a small time step, the gradient $\nabla h$ points in the right direction and is in line with the correct $\nabla d$ (but they do not have agreed magnitudes). Besides, there is a known fact about the geodesic function: $\nabla d$ always has unit magnitude. Then, by combining the direction given by $\nabla h$ and the magnitude governed by $\left\| \nabla d \right\| = 1$, the geodesic function can be attained.

The combination stated above can be implemented as the following three procedures. First, compute the heat function $h$ by solving the differential equation governing heat diffusion:
\begin{equation}\label{eq-diffusion}
	\frac{\partial}{\partial{t}}{h}(t, x) = {{\nabla }^{2}} h(t, x),
\end{equation}
where ${{\nabla }^{2}}$ is Laplacian operator. (The near-optimal total diffusion time is  the square of the mesh's mean edge length \cite{crane2017heat}.) Second, normalize the vector field $\nabla h$: 
\begin{equation}\label{eq-normalization}
 X = - \frac{\nabla h}{\left\| \nabla h \right\|}.
\end{equation}
Lastly, find the geodesic function $g$ that best fits $X$ by solving a least square problem:
\[ 
	\min_{g} \int_{\mathcal{M}} \left\| \nabla g - X \right\|^2,
\]
or equivalently, by solving its corresponding Euler-Lagrange equation:
\begin{equation}\label{eq-geodesic}
	{{\nabla }^{2}} g = \div (X),
\end{equation}
where $\div(\cdot)$ is the divergence operator. 

\begin{figure}[htbp]
	\centering
	\includegraphics[]{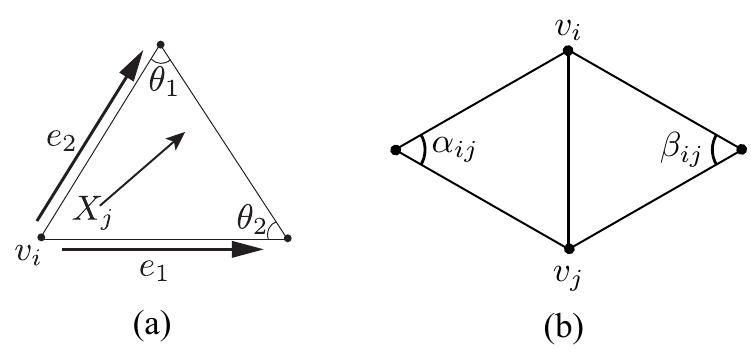}
	\caption{Triangles for computing divergence (a) and Laplaican (b).}
	\label{fig-Differential-Operators}
\end{figure}

Eq.~\ref{eq-diffusion}-\ref{eq-geodesic} involve three linear differential operators: $\nabla$, ${\nabla }^2$ and $ \div(\cdot)$. For mesh surfaces, their standard definitions are as follows (see \cite{botsch2010polygon} or the author's previous work \cite{zou2014iso} for detailed derivations). The gradient operator accepts a scalar function $f$, with $f_i$ at vertex $v_i$, and returns a 3D vector in each triangle $v_i, v_j, v_k$:
\begin{equation}\label{eq-gradient-operator}
	\nabla f = (f_j - f_i)\frac{n \times (v_i - v_k)}{2A} + (f_k - f_i)\frac{n \times (v_j - v_i)}{2A},
\end{equation}
where $n$ is the triangle normal and $A$ the triangle area. The divergence operator takes a vector field $X$ and gives back a scalar field with value at a mesh vertex $v_i$:
\begin{equation}\label{equ-div}
	\div(X)_i = \frac{1}{2A_i}\sum\limits_{k}\cot\theta_1(e_{1}^T X_j) + \cot\theta_{2}(e_{2}^T X_j),
\end{equation}
where $A_i$ is the Voronoi area of vertex $v_i$, the sum is taken over all incident triangles $k$ each with vector $X_j$, and $\theta_1, \theta_2, e_1, e_2$ are shown in Fig.~\ref{fig-Differential-Operators}a. The Laplacian of a scalar function $h$ gives another scalar function whose value at $v_i$ is given by:
\begin{equation}\label{equ-Laplacian}
	(\nabla ^2 h)_i = \frac{1}{2A_i}\sum\limits_{j} \left( \cot \alpha_{ij} + \cot \beta_{ij} \right)\left( h_j - h_i \right),
\end{equation}
where the sum is taken over all neighboring vertices $v_j$, and $\alpha_{ij}, \beta_{ij}$ are shown in Fig~\ref{fig-Differential-Operators}b.  It should be noted that the above schemes are sensitive to ill-shaped, skinny triangles. An additional preprocessing of local remeshing (by edge flipping and splitting) is thus carried out to avoid such situations.

With  (\ref{equ-div}) and (\ref{equ-Laplacian}), Eq.~\ref{eq-diffusion} and \ref{eq-geodesic} become two sparse systems of linear equations. For Eq.~\ref{eq-geodesic}, this is straightforward to see; for Eq.~\ref{eq-diffusion}, we need to use the backward difference to approximate the term $\frac{\partial}{\partial{t}}{h}$ and rearrange the equation into $\left( I - t \cdot \nabla ^2 \right)h(t) = h(0)$, where $I$ is the identity matrix, $t$ is the total diffusion time, and $h(0)$ denotes the initial heat distribution.

The last missing piece in converting geodesic computation to heat diffusion is to add the scallop metric $T_s$ to the above formulations. Loosely, a Riemannian metric like $T_s$ is a generalization of the first fundamental form of the design surface, which equips the surface with new inner products on the tangent planes \cite{Carmo76}. Inner products are responsible for calculating magnitudes of vectors and the angle between two vectors. So to add $T_s$ to the above formulations, we simply need to re-express all the calculations about distances and angles involved in Eq.~\ref{eq-diffusion}-\ref{equ-Laplacian} in terms of the scallop metric $T_s$. To be specific, the length of a vector $v$ is given by:
\[
	\| v \| = \sqrt{v^T T_s v}.
\]
The angle between two vectors $u, v$ is given by:
\[
	\theta = \arccos \left(\frac{u^T T_s v}{\| u \|\| v \|}\right).
\]
An equivalent but more compact way for adding the scallop metric is to first decompose $T_s$ into $T_s=D^T D$ and then apply the transformation 
\begin{equation}\label{equ-transformation}
	v^\prime = Dv
\end{equation} 
to each vector $v$ in question. The inner product on $v$ under the scallop metric is now equivalent to the usual inner product on $v^\prime$, as justified by the identity: $\|v^\prime\|_{Euclid}^2 = {v^\prime}^T v^\prime = (Dv)^T (Dv) = v^T (D^TD) v = v^T T_s v = \|v\|_{Scallop}^2$. Based on this transformation, all calculations in Eq.~\ref{eq-diffusion}-\ref{equ-Laplacian} remain unchanged except that vectors are transformed using Eq.~\ref{equ-transformation}. It should be noted that, as $T_s$ is positive definite, the decomposition $T_s=D^T D$ is always possible.

So far, the tool path generation problem has been formulated into the heat diffusion problem that mainly involves solving two systems of linear equations. Before closing this subsection, some discussions/insights about the robustness of the proposed method are provided. In the chain from tool path generation to geodesic computation to heat diffusion, the scallop metric is the bridging element that makes the chain possible, and the heat diffusion is the element contributing to the robustness and efficiency of the method. The diffusion step (Eq.~\ref{eq-diffusion}, in particular the Laplacian) is able to filter out the impact of mesh defects, in particular noises, due to its local averaging nature. For the same reason, the above elliptic PDEs can also be solved reliably \cite{coifman2006diffusion,crane2017heat}, although numerical approximation is in general sensitive to noises. (An exceptional mesh defect type that could degrade the method has been found to be ill-shaped, skinny triangles; local remeshing can help for such situations.) Although the diffusion step filters out mesh defects, it also makes tool paths deviate considerably from being iso-scallop. This is where the normalization step (Eq.~\ref{eq-normalization}) comes into play. It regularizes tool paths and enforces constant scallop heights. Altogether, they provide robustness towards generating iso-scallop tool paths.

Actually, the above discussion also implies an important limitation of the method. Sharp creases on the design surface are somewhat filtered out as well due to the local averaging in the diffusion step. As a result, generating iso-scallop tool paths near crease regions may not be supported satisfactorily. This limitation could be alleviated by segmenting the surface along creases and applying the method to individual segmented patches.

\subsection{The overall tool path generation method}
\label{sec:tool-path-algorithm}
The overall method consists of four steps:
\begin{enumerate}
	\item Construct the scallop metric $T_s$ for the mesh surface $M$ to be machined;
	\item Set up the initial tool path $C_0$ (typically, the surface boundary) by controlling the initial heat distribution $h(0): C_0 \rightarrow 1;  M \setminus C_0 \rightarrow 0$;
	\item Compute the geodesic function $g$ by solving Eq.~\ref{eq-diffusion}-\ref{eq-geodesic} and using Eq.~\ref{equ-div}-\ref{equ-transformation};
	\item Generate iso-scallop tool paths by extracting iso-level curves of $g$, with level increment $\sqrt{h}$ for a given scallop height $h$.
\end{enumerate}

In step 2, $h(0)$ enforces $C_0$ to be the source where heat spreads out over the rest of the surface, and then the $g$ computed in step 3 gives distances measured from $C_0$. $C_0$ is thus established as the initial tool path. In practice, the total diffusion time in step 3 can be changed instead of using the optimal value, e.g., using a larger time value for surfaces with severe noise. The introduced error duo to this change can be suppressed by, again, normalizing the gradient of $g$ and updating $g$ with this new gradient field. In step 4, the geodesic function $g$ can be reused to generate iso-scallop tool paths for different $h$, indicating the multiresolutional property of the proposed method.

%



\section{Results and discussion}
\label{sec:results}
The proposed method has been tested using models from both real and simulated data. Four case studies are to be presented to demonstrate the robustness and efficiency of the method. Case studies 1-3 are used to demonstrate the method's robustness feature, each responsible for one mesh defect type in noises, holes, and irregular density; case study 4 is a comprehensive case study involving all the defects. Case study 4 also presents comparisons with the classic iso-scallop tool path generation method by Feng and Li \cite{feng2002constant}. Case studies 1-3 were chosen to be simulated-data-based so as to have references/ground truths in analyzing errors of the generated tool paths. Case study 4 was based on real data. The efficiency of the method for the four case studies is summarized in Table \ref{performance-table}. The performance was measured based on a C++ implementation and a 2.4 GHz Intel Core i5 with 8G memory.

\subsection{Case studies}
\begin{figure*}[htbp]
	\centering
	\includegraphics[width = \textwidth]{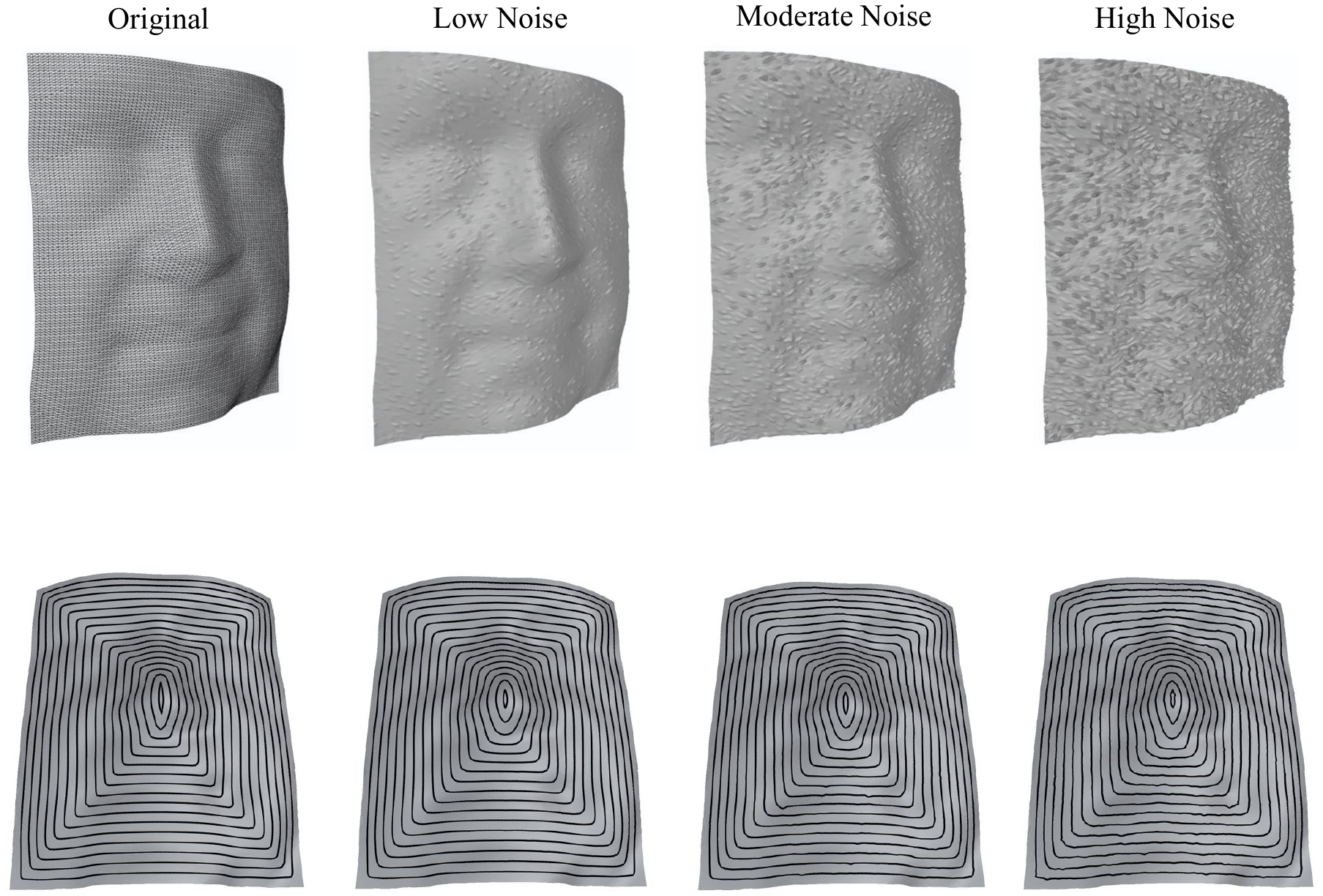}
    \caption{Tool path generation for a face model under various noise levels.}
	\label{face-results}
\end{figure*}

\begin{figure*}[htbp]
	\centering
	\includegraphics[]{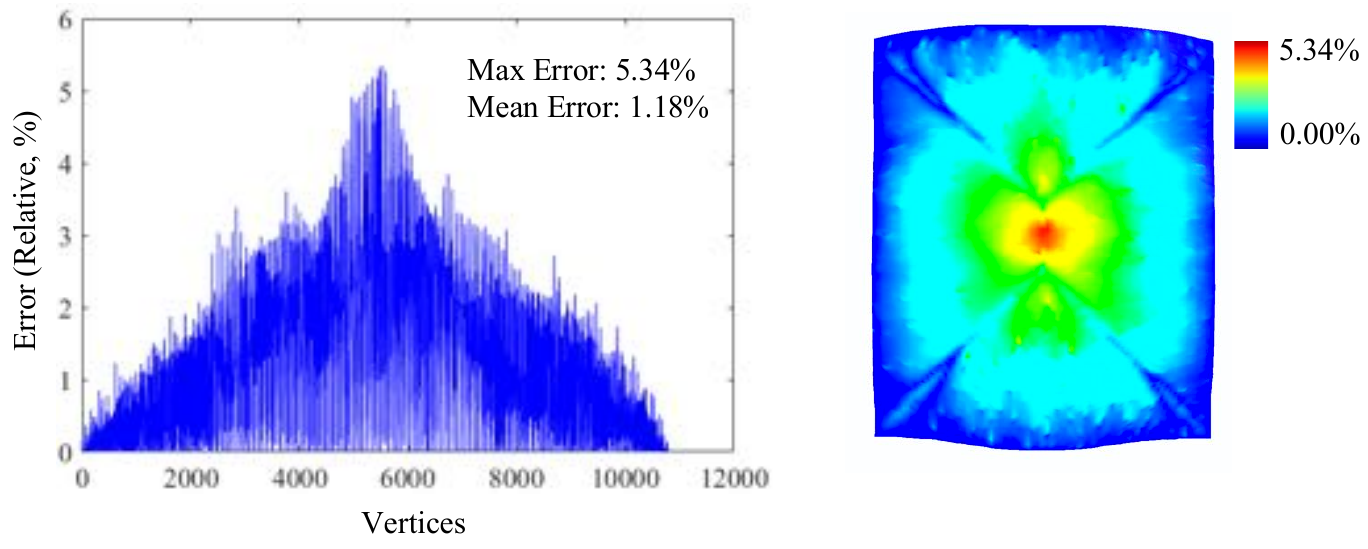}
    \caption{Tool path quality analysis for the face model.}
	\label{face-analysis}
\end{figure*}

Case study 1 considered a human face model where various levels of noises were applied (first row of Fig~\ref{face-results}). The original model is noise-free, and the figure underneath shows the iso-scallop tool paths generated for it. These tool paths served as references in analyzing errors of the tool paths generated for the noisy human faces. And the tool paths for the noisy faces are shown in the second row and besides the references (visualized on the noise-free face). A ball-end mill of radius $r = 4 mm$ was chosen, and the scallop height constraint was set as $0.1 mm$. The error analysis results shown in Fig.~\ref{face-analysis} were for the tool paths generated for the face with moderate noises. The error statistics are shown in the left figure, and the error distribution is given in the right figure. The error was measured relative to the mean side step.

\begin{figure*}[htbp]
	\centering
	\includegraphics[]{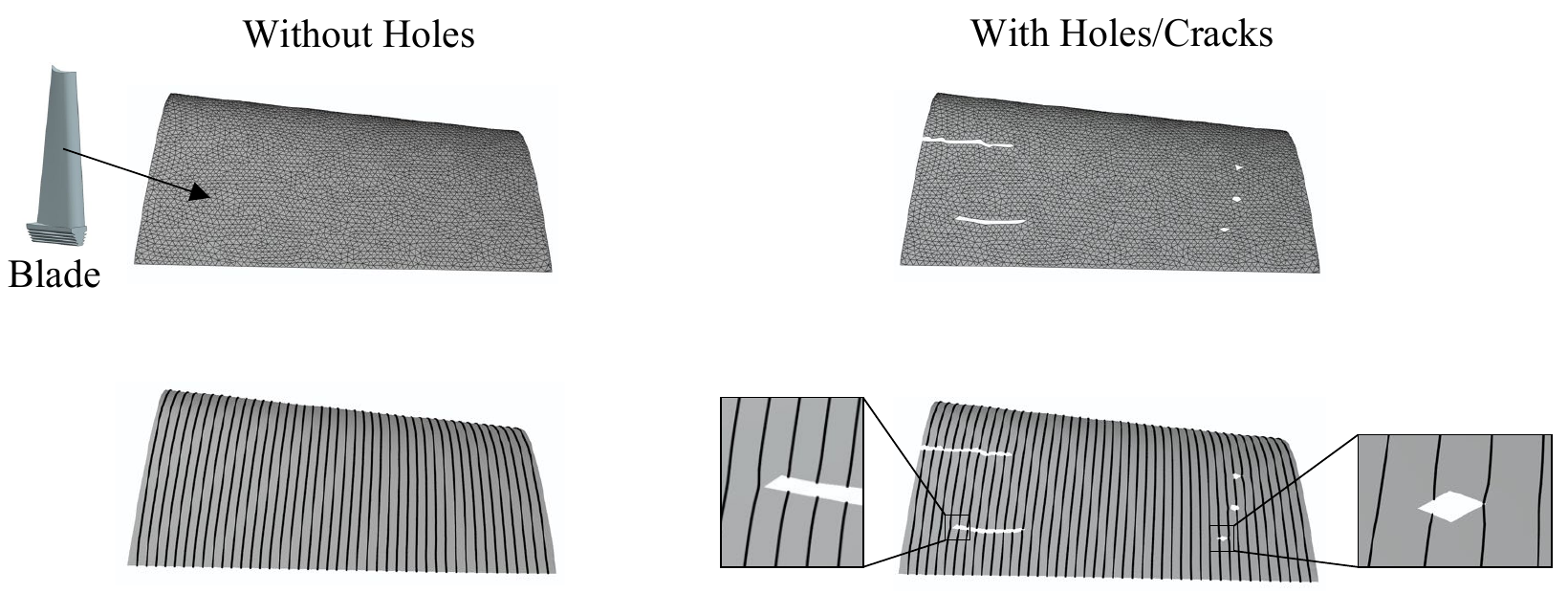}
	\caption{Tool path generation for a blade model with/without holes.}
	\label{blade-results}
\end{figure*}

\begin{figure*}[htbp]
	\centering
	\includegraphics[]{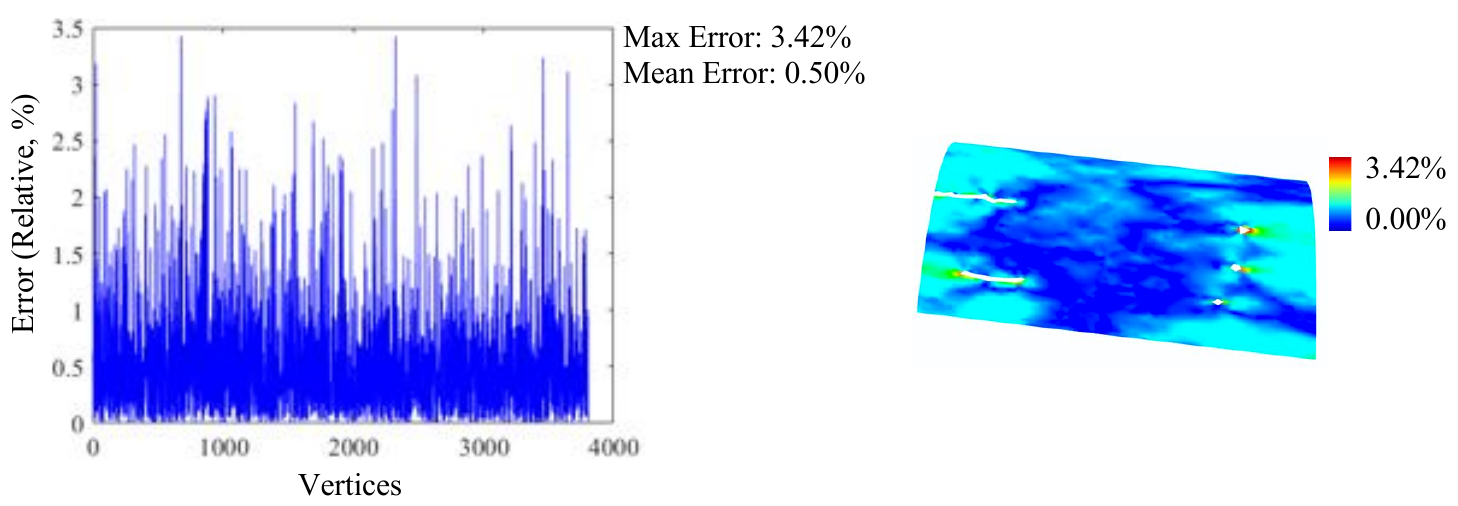}
    \caption{Tool path quality analysis for the blade model.}
	\label{blade-analysis}
\end{figure*}

Case study 2 involved a blade surface (Fig.~\ref{blade-results}), which may be the most widely used model in reverse engineering research \cite{su2020accurate}. Various types of holes were added to the surface, including open cracks, closed cracks, triangular holes, and round holes. The generated tool paths, with scallop height $1 mm$, for the original surface and the holed surface are shown in the second row; zoom-in views are also provided for better visualization of the holes' impact on generated tool paths. Error analysis for the generated tool paths is given in Fig.~\ref{blade-analysis}.

\begin{figure*}[htbp]
	\centering
	\includegraphics[]{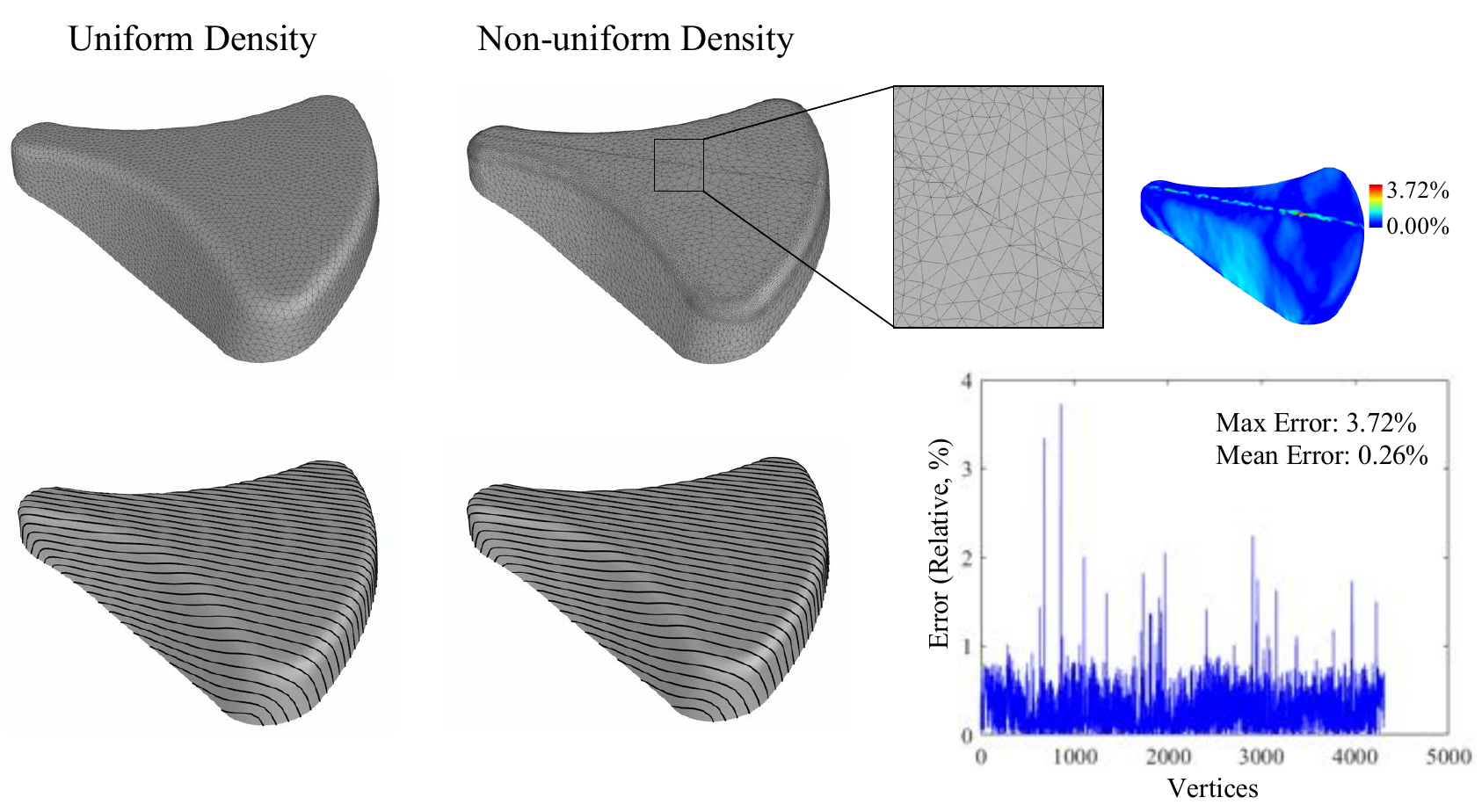}
	\caption{Tool path generation for a bicycle seat model under different density schemes and quality analysis.}
	\label{seat-results}
\end{figure*}

Case study 3 considered a bicycle seat model (Fig.~\ref{seat-results}) where two different meshing schemes were applied: uniform density and non-uniform density. The tool paths of the uniform model served as the reference model, and the tool paths of the non-uniform model demonstrated the impact of the variation of triangle density. Quantitative analysis of the impact was shown as the statistics and distribution figures besides the tool paths. The scallop height constraint was set as $0.1mm$.

\begin{figure*}[htbp]
	\centering
	\includegraphics[]{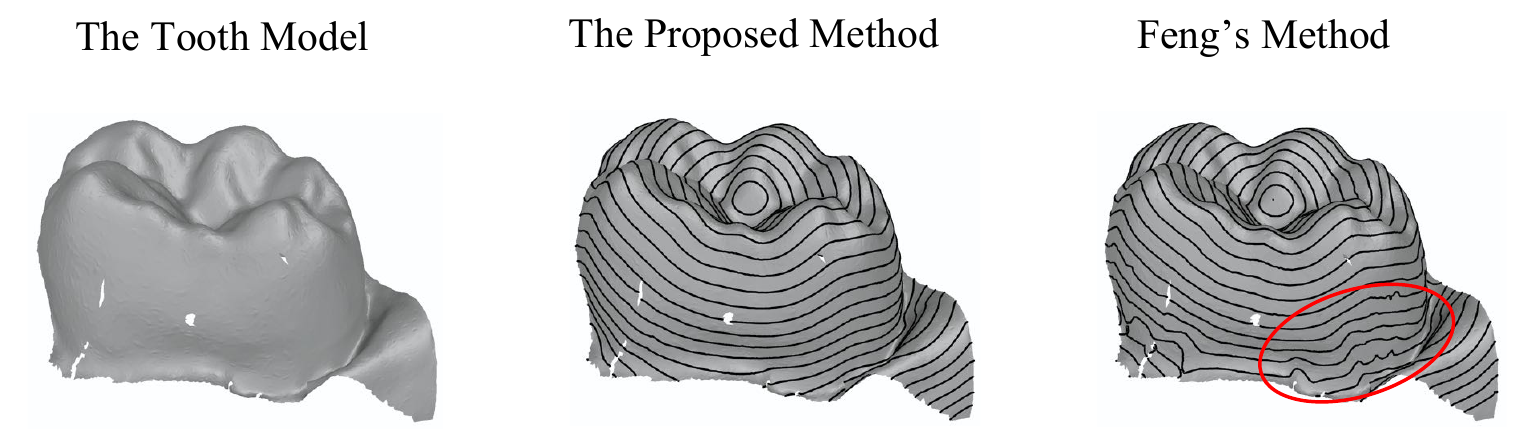}
	\caption{Tool path generation for a tooth model and comparisons with Feng's method \cite{feng2002constant}.}
	\label{comprehensive-results}
\end{figure*}

The case studies above deal respectively with the mesh defect types. The last case study was thus designed to contain a comprehensive dental model (Fig.~\ref{comprehensive-results}), which was downloaded from the GrabCAD part library (https://grabcad.com/library). This model possesses various mesh defects, and thus is a good candidate for showing the practical usage of the present work. The generated tool paths are shown in the middle figure of Fig.~\ref{comprehensive-results}. The scallop height constraint was set as $0.01mm$. An additional comparison with the classic iso-scallop method presented in \cite{feng2002constant} is provided to demonstrate the effectiveness of the proposed method. It should be noted that different scallop height constraints were used in the above case studies; this is because the models have different sizes, and adaptive constraints are needed for clear visualization of the generated tool paths.

\begin{table*}
	\renewcommand{\arraystretch}{1.25}
	\centering
	\caption{Performance of the proposed method in the case studies.}

	\begin{tabular}{cc|ccc} 
	\hline
	Models                                                               & Triangles                                 & Solving Time (s)                              & Tool Path Extraction Time (s)                & Total Time (s)  \\ 
	\hline
	\begin{tabular}[c]{@{}c@{}}Human Face\\(Moderate Noise)\end{tabular} & 21k & 0.176 & 0.016                                     & 0.192       \\
	\begin{tabular}[c]{@{}c@{}}Blade Surface\\(Holed)\end{tabular}       & 7.3k  & 0.059 & 0.009 & 0.068       \\
	\begin{tabular}[c]{@{}c@{}}Bicycle Seat\\(Non-uniform)\end{tabular}  & 16k & 0.122 & 0.015 & 0.137       \\
	Dental Tooth                                                         & 78k & 0.84  & 0.08                                      & 0.92    \\
	\hline
	\end{tabular}
	 \label{performance-table}
\end{table*}

\subsection{Discussion and limitations}

In all the above cases, the proposed method is seen to generate tool paths with very small errors, as expected. For the first three cases, the mean errors are consistently below $2\%$. According to Appendix B in \cite{Qiang2013iso}, the error on the side step leads to a doubled error on the scallop height. That being said, the mean errors of the scallop heights for the three cases are all below $4\%$. Thus, if the required scallop height constraint is $0.01 mm$, the proposed method can generate tool paths with a worst-case scallop height constraint around $0.01 \pm 0.0004 mm$, which can provide satisfactory precision control in general machining.

The comparisons shown in Fig.~\ref{comprehensive-results} further confirm the effectiveness of the proposed method. For the first few tool paths on the top face, Feng's method worked comparably well with the proposed method. Nevertheless, for the next tool paths on the side faces, Feng's method degraded significantly. The proposed method, on the other hand, continued to work well. The reason for these contrasting results is that the proposed method uses a global optimization (i.e., Eq.~\ref{eq-geodesic}) to obtain all tool paths at once, and this can evenly distribute error over the surface. Feng's method, as well as other existing methods, consistently use a sequential tool path generation approach that leads to accumulative errors in the tool paths. Due to this accumulation, these methods could generate good quality tool paths at the beginning but could soon become unstable and exhibit large errors in tool paths (e.g., as exemplified by those circled in Fig.~\ref{comprehensive-results}). For the same reason, we may conclude that the sequential tool path generation approach may be inherently unsuitable for the robust tool path generation problem.

By comparing the error analysis results in the first three case studies, it is also found that noises, holes, and irregularity are not of equal impact on tool path generation. Noise is the dominant factor in distorting the generated tool paths, holes have a moderate impact on the tool paths, and irregular density barely distorts tool paths except at some sparsely distributed locations (as shown by the error statistic figure in Fig.~\ref{seat-results}). As such, further improvements are better devoted to noise handling, and this seems to have a greater practical value.

It should, however, be noted that the above statements only hold for low or moderate levels of mesh defects. When high noise and/or large holes exist in the design surface, the proposed method should not be applied directly; instead, preprocessing such as denoising must be used in advance. Besides, as already noted, if the design surface has crease/sharp features, the proposed method could fail to generate iso-scallop tool paths near such regions. Although segmenting along the crease features may help, there are some surfaces that cannot be segmented satisfactorily. This is thus a serious limitation of the proposed method.

\section{Conclusions}
\label{sec:conclusion}
A new method has been presented in this paper to generate iso-scallop tool paths for triangular mesh surfaces. The main features of this method include the robustness against mesh defects such as noises, (small) holes/cracks, and irregular density, and the efficiency in computation. The robustness and efficiency are essentially achieved by converting the problem of tool path generation to the problem of heat diffusion, which in turn was broken down into two steps: (1) translate iso-scallop tool path generation to geodesic computation, and (2) convert geodesic computation to heat diffusion. New/improved methods have been presented to implement the conversion, and a series of case studies and comparisons have been carried to validate the methods.

A couple of limitations need to be noted here. During the implementation of the proposed method, it is found that the shape of triangles in the mesh surfaces can affect the robustness and run time of the proposed method. A successful application of the method requires that the mesh surface does not contain skinny triangles. Local remeshing could serve to address this issue. Another limitation is that the proposed method, in its current form, cannot deal with mesh surfaces involving crease features. This applicability limitation is inherent for the method, because the method's robustness feature is essentially attained through the local averaging nature of heat diffusion, which at the same time could average out sharp features. Segmentation techniques could alleviate this issue but may not be able to solve it completely. It should also be noted that the proposed method can effectively deal with noises, (small) holes/cracks, and irregular density, but other mesh defect types may not be supported satisfactorily. Extending the present work to addressing other mesh defect types can be very practically beneficial, and are among the CAM research studies to be carried out in the future.

\section{Acknowledgement}
\label{sec:7}
This work has been funded by a UBC PhD Fellowship and a grant from the National Key Basic Research Project of China.




\bibliographystyle{elsarticle-num}
\bibliography{bibliography}








%
%
\end{document}